\documentclass{aa}
\usepackage{natbib}
\usepackage{aasjournals}
\usepackage{xspace}
\usepackage{amssymb}
\usepackage{amsmath}
\usepackage{tabularx}
\usepackage{graphicx}
\bibpunct{(}{)}{;}{a}{}{,} 

\def\AU{\hbox{AU}}

\def\Fnir{{\ensuremath{F_{\mathrm{NIR}}}\xspace}}
\def\Ffir{{\ensuremath{F_{\mathrm{FIR}}}\xspace}}
\def\Fstar{\ensuremath{F_\star}\xspace}
\def\Fir{{\ensuremath{F_{\mathrm{IR}}}\xspace}}
\def\cdrev#1{\textbf{#1}}
\def\cdrev#1{#1}

\def\chiin{{\chi_{\mathrm{rim}}}}

\newcommand{\simless}{\mathbin{\lower 3pt\hbox
      {$\rlap{\raise 5pt\hbox{$\char'074$}}\mathchar"7218$}}} 
\newcommand{\simgreat}{\mathbin{\lower 3pt\hbox
     {$\rlap{\raise 5pt\hbox{$\char'076$}}\mathchar"7218$}}} 
\newcommand{\um}{{\ensuremath{\,\mu\mbox{m}}\xspace}}

\def\aap{A\& A}

\def\araa{AnnRevA\& A}

\def\apj{ApJ}
\def\aj{AJ}
\def\apjl{ApJL}
\def\apjs{ApJS}

\def\AU{\hbox{AU}}

\def\thttle{Understanding the spectra of isolated Herbig stars
  \newline in the frame of a passive disk model}
\begin{document}
\title{\thttle}
\author{C. Dominik\inst{1}, C.P. Dullemond\inst{2},
  L.B.F.M. Waters\inst{1,3}, S. Walch\inst{2}}
\authorrunning{Dominik, et al.}
\titlerunning{\thttle}
\institute{Sterrenkundig Instituut `Anton Pannekoek', Kruislaan 403,
  NL-1098 SJ Amsterdam, The Netherlands;\\ e--mail: dominik@science.uva.nl
\and
Max Planck Institut f\"ur Astrophysik, Karl
Schwarzschild Strasse 1, D--85748 Garching, Germany; e--mail:
dullemon@mpa-garching.mpg.de
\and
Instituut voor Sterrenkunde, Katholieke Universiteit Leuven,
Celestijnenlaan 200 B, B-3001 Heverlee, Belgium}
\date{Received 29 July 2002 / Accepted 8 November 2002}

\offprints{dominik@science.uva.nl.}

\abstract{We discuss spectral energy distributions of a sample of
  Herbig Ae/Be stars in the context of a passive irradiated disk
  model. The data have been presented earlier by
  \citet{2001A&A...365..476M}, and preliminary interpretations of
  these data were given in that paper.  While the spectra of Herbig Ae
  stars all show similarities, there is significant variation between
  the spectra, in particular in the shape of the mid-IR rise and in
  the presence or absence of a silicate feature.  We explore the
  hypothesis that all these different spectra can be interpreted as
  pure disk spectra without additional components. Using the model of
  Dullemond, Dominik and Natta (\citeyear{2001ApJ...560..957D}) we
  deduce the disk parameters of a number of the sources, and find that
  for a large fraction of investigated sources, satisfactory fits can
  be obtained.  The derived model parameters show that some group Ia
  sources can only be fit with radially increasing surface densities,
  indicating the presence of depleted inner disk regions.  The
  steep-sloped SEDs of group IIa sources can be fit with very compact
  disks, probably representing disks with collapsed outer regions.
  The largest difficulties arise from sources that do not show
  significant silicate emission features.  Our attempts to explain
  these objects with a pure geometric effect are only partially
  successful.  It seems that these stars indeed require a strong
  depletion of small silicate grains.}

\maketitle


\section{Introduction}

\cdrev{Herbig Ae/Be stars (herafter referred to as HAEBE stars) are
  thought to be young stellar objects of intermediate mass
  \citep{1960ApJS....4..337H,1972ApJ...173..353S,1984A&AS...55..109F,1994A&AS..104..315T,1998A&A...331..211M}.
  They are associated with large amounts of circumstellar matter, both
  gas and dust, and are found in or near star forming regions. In the
  seminal paper of \citet{1960ApJS....4..337H}, HAEBEs were defined to
  be near a reflection nebulosity, because this ensured they are young
  and close to a star forming region.  The close proximity to
  molecular cloud material in combination with limited angular
  resolution observations makes it hard to distinguish dust and gas
  emission from the loose surroundings from that of the circumstellar
  disk. However, objects have been found that are very
  similar to other HAEBEs but lack a reflection nebulosity \citep[see
  e.g.][]{1998ARA&A..36..233W}.  These stars are called \emph{isolated
    Herbig stars}.  Hipparcos observations of the nearest HAEBE stars
  have confirmed their pre-main-sequence nature
  \citep{1997A&A...324L..33V}.}
  
\cdrev{HAEBE stars have a strong infrared (IR) excess due to
  circumstellar dust, often carrying a considerable fraction of the
  total luminosity of the system. Mid-IR imaging has shown that some
  HAEBE stars have spatially extended emission at these wavelengths
  \citep{1994A&A...292..593P,1994ApJ...432..710D,1998ApJ...509..324D};
  these stars often (but not always) are of (early) B spectral type,
  and their emission can be understood in terms of heating of the
  surrounding molecular cloud by the strong UV radiation field of the
  young star, possibly in combination with a disk
  \citep{1999ApJ...520L.115M,1998A&A...336..565H}.  However, the less
  luminous late B, and A type stars in the HAEBE class often show
  unresolved or compact emission in the mid-IR when observed with 4
  meter class telescopes on a scale of 1--2 arcsec, corresponding to
  100-200 AU at typical distances of 100\,pc.  Van den Ancker et al
  (private communication) have unsuccessfully tried to resolve several
  of the sources in the sample (HD\,100546, HD\,104237, HD\,144432,
  HD\,163296) with the ESO 3.6m telescope, finding only upper limit of
  typically 0.7''.  Other observations have succeeded resolving the
  emission of some stars (e.g. AB Aurigae \citep{1995ApJ...451..777M},
  HD100546 \citep{2001AJ....122.3396G}, HD97048 (van Boeckel et al, in
  preparation)) and found compact sizes.   The isolated Herbig stars 
  usually show little or no reddening at optical wavelengths, which is
  remarkable given their large L$_{\rm IR}$/L$_{*}$ ratio.  An obvious
  way to explain this observation is to assume a flattened, disk-like
  distribution of the gas and dust viewed at some intermediate
  inclination angle.  This disk can be understood as the passively
  heated remnant of the accretion disk which was present during the
  initial phase of star formation, and is similar to disks seen around
  the lower mass T~Tauri stars. Since such passive disks are believed
  to be the site of planet formation, HAEBE stars have gained
  considerable attention in recent years.}
  
\cdrev{Obviously, the best way to determine the geometry of the
  circumstellar material around isolated HAEBE stars is by direct
  imaging.  \citet{1997ApJ...490..792M} spatially resolved the
  millimeter continuum and line emission from the Herbig Ae star
  HD163296 and found an elongated structure. Submillimeter aperture
  synthesis images in CO lines show rotation profiles in a number of
  Herbig Ae stars, with rotational velocities consistent with a
  Keplerian disk
  \citep{1998A&A...338L..63D,1999alma.confE..50Q,1997ApJ...490..792M,2000ApJ...529..391M}.
  Based on this evidence one may conclude with some confidence that at
  least the outer parts of the circumstellar matter distribution
  around some (isolated) Herbig Ae stars are, just as in the case of
  T-Tauri stars, in fact rotating circumstellar disks.}
  
\cdrev{Observations of the spatial distribution of the circumstellar matter
  closer to the star (inwards of about 100 AU) have not yet resulted
  in a clear picture. There seems to be conflicting evidence
  suggesting both a flattened and more spherical distribution of gas
  and dust. Near-IR and optical imaging using ground-based adaptive
  optics and HST in several cases shows the emission to be flattened
  \citep{2000ApJ...544..895G,2001AAS...198.7716G,2001AJ....122.3396G,2000A&A...361L...9P}.
  However, interferometric observations at 2 $\mu$m, probing the inner
  regions on scales below 10 AU, have shown little or no evidence for
  a disk-like geometry \citep{2001ApJ...546..358M}.}

In view of the scarcity of resolved observations at these spatial
scales, it seems that the procedure of fitting a model to the observed
SED is still a valuable tool to study the properties of circumstellar
matter around young stars. Under the assumption that the circumstellar
dust is in radiative equilibrium, the temperature of the dust can be
calculated by solving the radiative transfer equation. By varying the
density distribution and the dust composition, one can eventually find
a good 'fit' to the data.  One has to keep in mind however that this
procedure can not uniquely determine the geometry and density
distribution of the circumstellar matter. Often, a multitude of
different parameter sets lead to equally good fits
\citep{1994A&A...287..493T}. This clearly means that the SED alone
does not contain enough information about the geometry and dust
distribution in these objects.

Nevertheless, model fitting can retrieve interesting information from
the SED if one uses additional constraints to the density
distribution. These constraints can be obtained from
e.g.~spectroscopic quantities (lines, features), or any kind of
imaging data when available. In addition to this, one can add {\em
  physics} to the model. For instance, one could require the density
distribution to be in hydrostatic equilibrium: a self-consistent disk
model. A simple, but powerful model of this kind is the semi-analytic
passive flaring disk model of \citet{1997ApJ...490..368C}. This model
naturally reproduces the shape of the spectrum of many Herbig Ae stars
longwards of 6 micron. Moreover, as was first noted by
\citet{2001A&A...371..186N}, if one correctly accounts for emission
from the disk's inner rim, also the SED shortwards of 6 microns (the
conspicuous 3 micron bump) can be explained. In a recent paper,
Dullemond, Dominik \& Natta (2001, henceforth DDN), presented such a
self-consistent disk model including both the emission from the
flaring part of the disk and from the inner rim. They found that
emission from the hot inner rim (being hotter than the flaring disk
and thus having a higher scale height) can account for the near-IR
excess in the HAEBE star AB~Aur; the DDN model for AB~Aur is the first
disk-only model that can account for the \emph{entire} SED of AB~Aur.

Encouraged by our success in fitting AB~Aur, we decided to investigage
the applicability of the DDN model to a wider sample of HAEBE stars.
\cdrev{We use the sample described by \citet{2001A&A...365..476M}.
  This is a set of isolated Herbig Ae stars which was selected from a
  larger set of Herbig AeBe stars
  \citep{1994nesh.conf..405W,1994PhDT.......226B,1998A&A...331..211M}.
  To our knowledge, none of the stars (with the exception of AB~Aur)
  shows any significant reflection nebulosity or significant extended
  emission on scales much larger than expected for a disk.  This
  suggests that the SED is not contaminated by large scale dust clouds
  surrounding the object.  The objects were further selected by their
  very low circumstellar extinction - typically 0.5 magnitudes or
  less.  Exceptions are HD\,142666 (0.9\,mag) and HD\,150193 (1.5\,mag)
  \citep[and private communication]{1998A&A...330..145V}}.  In their
paper, Meeus et al.~presented combined ISO SWS/LWS spectra and
literature photometry data of 14 Herbig Ae/Be stars, and found that
these objects show both striking similarities in the spectrum, and
important differences.

We will attempt to interpret this variety of spectra entirely in the context
of the DDN model, without additional components. This way we can find out
whether an isolated disk picture is sufficient to explain the SEDs of these
stars. And if we can, we will see what we can learn from the deduced
parameters.  The paper is organized as follows.  In section
\ref{sec:sample-seds} we describe the sample.  In
section~\ref{sec:model-fitt-proc} the DDN model is introduced and we
detail the fitting procedure.  In section~\ref{sec:results}, the
obtained fits are shown and discussed.  Finally, in
section~\ref{sec:discussion}, we reflect on the successes and failures
of the model to reproduce the SEDs of isolated Herbig stars.

\section{Sample and SEDs}
\label{sec:sample-seds}
We use the sample of isolated Herbig stars as given by Meeus et al.
This is a group of 13 stars which have the properties of Herbig Ae
stars, but which are not located very closely to star forming regions.
The original sample contains 14 stars, but we removed 51 Oph because
of its uncertain nature \citep{2001A&A...369L..17V}).  The remaining
SEDs are shown in figs.~\ref{fig:groupIa}--\ref{fig:groupIIa}.
%
%
The
source of the IR excess in these stars is believed to originate from a
disk.  The SEDs of these sources is similar, but still shows
considerable variation.  All stars show the characteristic bump in the
spectrum around 3 micron.  At a wavelength of about $5\um$, the
SEDs ($\nu F_{\nu}$) show a local minimum, followed by the 10\um{}
region in which most stars show a prominent emission feature
attributed to silicate.  Longwards of 10\um{}, the SEDs differ
significantly. Meeus et al.~found that the SEDs can be classified into
two major groups, group I and group II, and they speculate that the
difference between the groups can be attributed to geometry.  Group I
sources are objects in which the SED plotted as $\nu F_{\nu}$ stays
high or even rises in the region between 20\um{} and about 100\um{}.
Meeus et al.~attribute this strong mid-IR component to a flaring disk.
Group II sources on the other hand continue to decrease longwards of
20\um{} (the ``powerlaw component'' in Meeus et al), their spectrum in
this wavelength region is more consistent with that of a
geometrically thin disk.  Numerically, the distinction between groups I and
II can be measured by the different fractions of stellar light emitted
in the near IR (NIR) and far IR (FIR) region. Table~\ref{tab:measured}
shows these quantities for all stars in the sample.  The stellar flux
\Fstar is measured by integrating the Kurucz model fit to the optical
and UV spectrum.  In order to compute the emission from the disk, we
subtract the Kurucz model from the SED and integrate the excess flux
from 2\um{} to 7\um{} to define \Fnir{} and from 7\um{} to
infinity to define \Ffir.  In the NIR, almost all stars emit between
10 and 30\%, with significant variation from source to source.  In the
FIR, group I sources reprocess typically 20-30\% of the stellar
radiation, while the group II sources are closer to 15\%.  The total
effect of NIR and FIR reprocessing can be seen in the quantity
\Fir/\Fstar which shows that group I sources reprocess in total around
50\% of the stellar radiation, group II sources only between 20 and
30\%.

\begin{table}
\caption{\label{tab:measured}Measured properties}
\begin{tabular}[tb]{cllccccc}
Group& Name       &  $\frac{\Fnir}{\Fstar}$ & $\frac{\Ffir}{\Fstar}$ &
  $\frac{\Fir}{\Fstar}$ & $\frac{\Fnir}{\Fir}$ &
  $\frac{F_{{10\mu\mathrm{m}}}}{F_{{7.7\mu\mathrm{m}}}}$ \\\hline
Ia   & AB Aur     &  0.12   & 0.22 & 0.42   & 0.47  &   1.56\\
Ia   & HD100546   &  0.07   & 0.39 & 0.45   & 0.15  &   6.95\\
Ia   & HD142527   &  0.32   & 0.60 & 0.92   & 0.35  &   1.33\\
Ia   & HD179218   &  0.10   & 0.24 & 0.35   & 0.30  &   3.85\\
\hline
Ib   & HD100453   &  0.22   & 0.29 & 0.50   & 0.43  &   2.56\\
Ib   & HD135344   &  0.24   & 0.29 & 0.53   & 0.46  &   1.10\\
Ib   & HD139614   &  0.09   & 0.23 & 0.32   & 0.28  &   6.53\\
Ib   & HD169142   &  0.04   & 0.07 & 0.11   & 0.35  &   4.19\\
\hline
IIa  & HD104237   &  0.16   & 0.08 & 0.24   & 0.66  &   1.17\\
IIa  & HD142666   &  0.14   & 0.15 & 0.29   & 0.47  &   1.82\\
IIa  & HD144432   &  0.13   & 0.13 & 0.27   & 0.50  &   3.28\\
IIa  & HD150193   &  0.14   & 0.13 & 0.27   & 0.52  &   1.53\\
IIa  & HD163296   &  0.21   & 0.17 & 0.38   & 0.54  &   1.53\\
\hline
\end{tabular}
\end{table}

Meeus et al.~also introduced a subdivision of the groups into the
subgroups a and b, which differ in the 10\um{} region.  Sources in the
``a'' subgroup show strong silicate emission, while sources in the
``b'' subgroup have no detectable silicate emission at all.  While the
signal-to-noise ratio in some of the group ``b'' sources is limited,
there is at least one case (HD\,100453) where the evidence for the
suppression of the 10\um{} feature is very strong.  Interestingly, all
``b'' sources in the Meeus et al.~sample belong to group I, while in
group II all stars show prominent silicate emission.

\section{Model and fitting procedure}
\label{sec:model-fitt-proc}
\subsection{The DDN model}
\begin{figure}[t!]
\includegraphics[width=9cm]{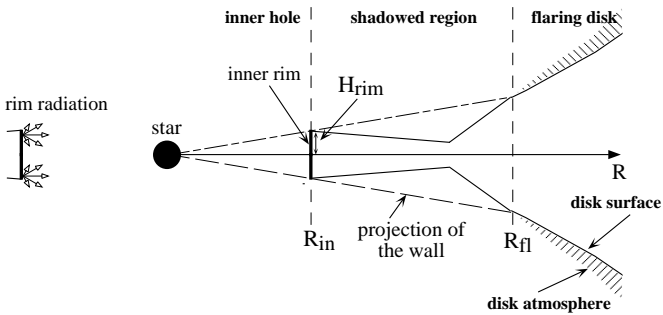}
\caption{\cdrev{Sketch of the geometrical structure of the DDN model.  See
  text for details.  After \citet{2001ApJ...560..957D}.}}
\label{fig-sketch}
\end{figure}

The DDN disk model is based on the flaring passive disk model of
\citet{1997ApJ...490..368C}.  Improvements were made to that original
model in order to conserve energy \citep[see
also][]{2001ApJ...547.1077C} and to properly account for the reduction
of irradiation flux due to occultation of part of the stellar surface
by the disk itself.  Most importantly, the DDN model includes a proper
treatment of the inner rim, which is located at the dust evaporation
radius. This radius is computed self-consistently.  The DDN model
assumes that the material inside the dust evaporation surface is
optically thin to the stellar radiation.  The unattenuated stellar
flux impinges on the inner edge of the dusty part of the disk. Since
the the inner rim exposes a vertical surface on which the stellar flux
impinges perpendicularly, it gets much hotter than expected for a
flaring disk at the same radius: the inner rim puffs up.  The DDN
model computes how much the inner rim puffs up, and how much of the
disk behind it will be shadowed by this puffed-up rim \cdrev{(see
  fig.~\ref{fig-sketch}}). 
Effects of self-irradiation of the disk are included, as well as a
simple treatment of radial radiative diffusion in the disk midplane,
which can be important for the structure of the shadowed region. The
disk model can treat low optical depths, as long as the vertical
surface height is large enough that a $\tau=1$ surface can be defined
for the direct stellar photons impinging under a grazing angle onto
the disk.  For very low mass disks, or the very outer parts of
intermediate mass disks, in which even the radial optical depth at
stellar wavelengths is below unity, the DDN approach breaks down.

The puffed-up inner rim is responsible for the 3-micron bump in the
spectrum \citep{2001A&A...371..186N}.  The vertical 'wall' at 1500 K
produces the largest flux under an inclination $i$ close to edge-on.
However, if the inclination is too large (close to edge-on), the
observers sees the inner regions of the disk through the outer disk,
and the inner rim as well as the central star will be obscured.  None
of the stars in the Meeus sample have large extinction towards the
star, so we can rule out this extreme case.  In the DDN model, the 3
micron flux also vanishes for exact face-on inclination ($i=0$). This
is a geometrical effect and is due to the fact that the emissions
originates from the surface of a cylinder.  The projected surface of
this cylinder vanishes for low inclinations.  In reality, the wall
will likely be curved, weakening this effect.

The disk directly behind the inner wall is shadowed by the wall
itself.  It does not receive direct stellar radiation and will
therefore not radiate significantly.  In the SED, the contribution of
this region is negligible.  At larger radii, the flaring disk
reappears from the shadow, and emits as the usual flaring passive
irradiated disk, producing mid- and far infrared emission. If the disk
is rather massive, it will be optically thick at most wavelengths in
the infrared and produce a rather flat spectrum (in $\nu
F_\nu$).  The flat spectrum which is a typical sign of flaring disks
results from the sum of two equally strong components: an optically
thin component in the mid-IR with emission features (the disk's
surface layer), and an optically thick component at far-IR wavelengths
(the disk's interior).

\begin{figure}[b!]
\includegraphics[width=9cm]{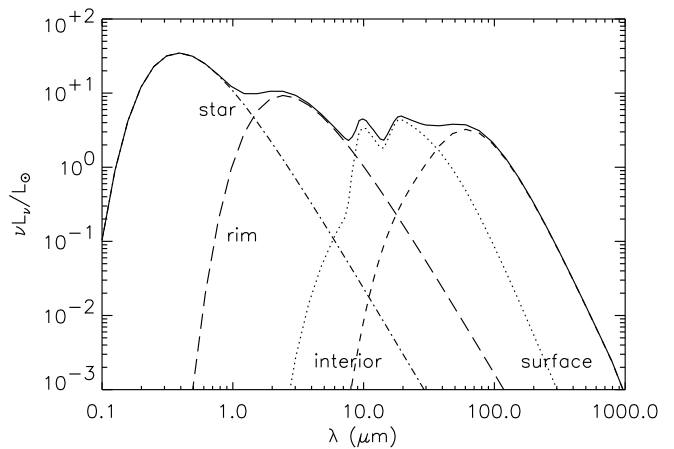}
\caption{An example of an SED of a Herbig Ae star, as computed by the
DDN model. The spectrum as shown here is decomposed in its four main
components: star, inner rim, disk surface and disk interior. The star
is approximated with a blackbody curve.}
\label{fig-example}
\end{figure}

The spectrum of the disk will therefore consist of three major components:
the 3 micron emission from the inner rim, the mid-IR emission from the
surface layer, and the far-IR emission from the disk's interior (see
Fig.\ref{fig-example}).
The basic shape of the SED can be easily understood in terms of energy
conservation. The total flux in the infrared excess compared to the
stellar flux is determined by the covering fraction of the disk with
respect to the central star. This covering fraction is equal to the
ratio $H_s/R$ (surface height over radius) at the outer edge of the
disk, where the disk geometric thickness is maximum. In general one
can say that the covering fraction increases for larger mass and/or
larger outer radius, and is only weakly dependent on other parameters
of the disk. The other parameters determine how this fixed amount of
IR excess is divided over the three components. For instance, a large
inner rim height will put a lot of flux in the 3-micron bump at the
cost of the mid- and far-IR. For more details on how the disk
parameters affect the SED we refer to the DDN paper.

\subsection{The dust opacities}

In the disks we are studying, the continuum opacity is entirely
dominated by the dust. We use dust grains of 0.1 micron in size,
belonging to one of the following different dust species: astronomical
silicate \citep{1984ApJ...285...89D}, carbon
\citep{1993ApJ...402..441L}, and forsterite
\citep{servoin,1996ApJS..105..401S}.  These three species can be mixed
in any composition, but they are assumed to be thermally coupled,
i.e.~we assume a single temperature for all dust species. In most
cases we assume a standard mixture of dust: 90\% silicate and 10\%
carbon.  We do not attempt to fit in detail all the spectral features
in the SED, in particular we do not include PAH's even though they are
clearly present in some objetcs.  However, in one case (HD 100546) we
include crystalline forsterite in order to match the extremely strong
contribution of these features in this source.  In general, however,
we focus on fitting the SED which is only weakly dependent on the
actual dust composition. The exception is the (sub-)mm range of the
spectrum: this part is critically dependent on the long-wavelength
behavior of the silicate opacity. There is long standing debate about
the slope of the opacity in this range. The opacity of Draine \& Lee
goes as $\kappa_\lambda\propto \lambda^{-2}$ while from observations
of the interstellar medium it follows that $\kappa_\lambda\propto
\lambda^{-1}$ \citep{2000prpl.conf..533B}. Many of
the sources in our sample (with the notable exception of AB Aurigae)
can be fitted better with an opacity that goes as
$\kappa_\lambda\propto \lambda^{-1}$ at long wavelengths. So we add an
adapted silicate opacity that is identical to the Draine \& Lee
opacity for $\lambda<100\mu$m and goes as $\kappa_\lambda\propto
\lambda^{-1}$ for $\lambda>100\mu$m. By gluing the two opacity regimes
smoothly together at $\lambda=100\mu$m, the long wavelength opacity in
fact turns out to have exactly the same normalization as the opacity
proposed by \citet{2000prpl.conf..533B}.


\noindent
\begin{table}
\caption{\label{tab:ddnparameters} The parameters entering into the
  DDN model.}
\begin{tabular}{lp{5.5cm}}
Parameter         & observed or fitted? \\\hline
$L_{\star}$       & from observations (Meeus et al.)\\
$M_{\star}$       & from observations (Meeus et al.)\\
$T_{star}$        & from observations (Meeus et al.)\\
T$_{\rm rim}$     & fixed at 1500K \\
Dust              & mostly fixed composition: 0.1\um{} grains, 5 or 10\%
carbon, the rest silicate.  $\beta$ slope of silicate opacity either
one or two, fitted.\\
M$_{\rm disk}$    & fitted (minimum value if not unique) \\
$p$               & fitted\\
$r_{\rm out}$     & fitted\\
$\chi$            & fitted for group Ib sources, otherwise very close
                    to selfconsistent value.\\
$i$               & inclination from observations when available, otherwise
                    fitted\\
\end{tabular}
\end{table}

\subsection{Fitting procedure}
In order to fit the SEDs of the stars, we run the DDN model and change
parameters until we find a fit to the observed spectrum. To minimize
the chance of missing important regions of parameter space, and to
maximize our understanding of the fitted solutions, we employ a
specially developed a widget interface that allows one to do
on-the-spot fitting of the model to the observed SED.

In our fitting procedure we do not vary all the parameters of the DDN
model.  Whenever possible, we used observations to constrain specific
parameters.  In particular, the stellar properties were inferred from
observed parallax, magnitude and spectral type. There are some
uncertainties in the inferred values of, in particular, the mass and
the luminosity of the star.  We have taken the luminosities provided
by Meeus et al, and the masses given by \citet{1997A&A...324L..33V}.
The uncertainties in the luminosity stem from the error in the
Hipparcos distances and from uncertainties regarding the dereddening
which is crucial for the determination of the bolometric luminosity of
hot stars.  The uncertainties in the luminosities are typically up to
50\%.  The stellar masses were obtained by placing the stars into an
HR diagram with overlayed pre-main-sequence evolutionary tracks.  The
uncertainties in the masses are relatively small. Van den Ancker et al
quote 5-10\% statistical error.

The main disk parameters that are varied in order to fit the SED are
the disk mass $M_{\rm disk}$, the slope $p$ of the powerlaw describing
the radial dependence of the surface density, the outer disk radius
$r_{\rm out}$, the inclination $i$ and, in some cases, the height of
the inner rim in units of the local pressure scale height $\chi_{\rm
  rim}\equiv (H_s/H_p)_{\mathrm{rim}}$. The latter parameter is
usually computed self-consistently, but occasionally is 
fine-tune by hand.  For some sources (AB Aur, HD 100546, HD 163296)
the outer radius could be assumed to be the radius measured by
continuum interferometry at 2.6\,mm \citep{1997ApJ...490..792M}.
However, we consider these radii to be inaccurate, since they only
tell at which radius the signal-per-synthesized-beam drops below the
noise. We therefore kept the outer radius a free fitting parameter.
The observed radii do, however, provide a lower limit to the disk
radius.

When we try to fit a particular SED, we put most of the weight on the
IR region of the spectrum.  This is the region where most of the
reprocessed stellar energy emerges, and therefore this part of the
spectrum is most strongly constrained by the model. This approach
sometimes leads to underestimating the submm fluxes.  While not
completely satisfactory, the deviating submm fluxes are energetically
irrelevant.  The large uncertainties in the mm and radio dust
opacities also indicate that these points provide less solid
constraints to disk models.  Additional flux in the submm region can
be produced by introducing an addtional component of large dust grains
in the midplane, as has been done by many authors
\citep[e.g.][]{2001A&A...371..186N,2000A&A...360..213B}.  This is
basically equivalent to changing the wavelengths dependence of the
opacity in the region.

Table \ref{tab:ddnparameters}  summarizes the input parameters of a DDN model.

\begin{table*}
\caption{\label{tab:parameters}Stellar properties and fit parameters. In the
column of $\chi$ the numbers which are bracketed are results of the
self-consistent determination of this parameter. If not bracketed, it is
fitted by hand.}
\begin{tabular}[tb]{cllrrrr|rrrrrrrrr}
Group & Name & SpType & $M_{\star}$ & $L_{\star}$ & $T_{\rm eff}$ & $d$ & $M_{\rm disk}$ & $p$ & $\chi$ & $r_{\rm out}$ & $i$ & $f_{\rm sil}$ & $f_{\rm sil1}$ & $f_{\rm fors}$ & $f_{\rm c}$ \\
      &      &     & $M_{\odot}$ & $L_{\odot}$ & K             & pc  & $M_{\odot}$    &     &        &  AU           &               &                &                &             &   \\\hline
 Ia &   AB Aur &      B9/A0Ve &  2.50 &    47 &  9750 & 144 &  0.100 &   -2.0 &     (5.5) &  400 & 65 & 95 &  0 &  0 &  5 \\
 Ia & HD100546 &      B9Ve    &  2.50 &    36 & 11000 & 103 &  0.005 &    0.0 &       1.2 &  400 & 51 &  0 & 70 & 17 & 11 \\
 Ia & HD100546 &      B9Ve    &  2.50 &    36 & 11000 & 103 &  0.020 &    2.0 &     (0.0) &   43 & 30 & 47 &  0 & 47 &  5 \\
 Ia & HD142527 &      F7IIIe  &  2.50 &    31 &  6250 & 200 &  0.010 &   -1.0 &     (4.3) &  200 & 75 &  0 & 90 &  0 & 10 \\
 Ia & HD179218 &      B9e     &  2.70 &    80 & 10000 & 240 &  0.010 &    0.8 &     (3.4) &   30 & 20 & 80 &  0 &  0 & 20 \\
\hline
 Ib & HD100453 &      A9Ve    &  1.70 &     9 &  7500 & 114 &  0.010 &   -0.5 &       6.0 &  600 & 65 & 90 &  0 &  0 & 10 \\
 Ib & HD100453 &      A9Ve    &  1.70 &     9 &  7500 & 114 &  2.000 &   -1.0 &     (5.3) &  300 & 75 &  0 &  0 &  0 & 10 \\
 Ib & HD135344 &      F4Ve    &  1.30 &     3 &  6750 &  84 &  0.010 &   -0.8 &       8.0 &  800 & 60 &  0 & 90 &  0 & 10 \\
 Ib & HD139614 &      A7Ve    &  1.80 &    12 &  8000 & 157 &  0.010 &   -0.1 &       5.0 &   60 & 20 &  0 & 94 &  0 &  5 \\
 Ib & HD169142 &      A5Ve    &  2.50 &    32 & 10500 & 145 &  0.100 &   -2.0 &       7.0 &  100 &  8 &  0 & 90 &  0 & 10 \\
\hline
IIa & HD104237 &      A4Ve    &  2.60 &    40 & 10500 & 116 &  0.060 &    0.0 &     (4.8) &   10 & 55 &  0 & 90 &  0 & 10 \\
IIa & HD142666 &      A8Ve    &  1.80 &    11 &  8500 & 116 &  0.030 &    0.5 &     (4.3) &   10 & 55 &  0 & 90 &  0 & 10 \\
IIa & HD144432 &      A9Ve    &  2.20 &    32 &  8000 & 200 &  0.200 &    0.5 &     (4.8) &   10 & 45 &  0 & 90 &  0 & 10 \\
IIa & HD150193 &      A1Ve    &  2.50 &    40 & 10000 & 150 &  0.010 &    0.5 &     (4.3) &    8 & 45 &  0 & 90 &  0 & 10 \\
IIa & HD163296 &      A3Ve    &  2.40 &    30 & 10500 & 122 &  0.050 &   -0.2 &     (4.3) &   50 & 65 &  0 & 90 &  0 & 10 \\
\hline

\end{tabular}
\end{table*}

\begin{figure*}[t!]
\includegraphics[height=22cm,width=19cm]{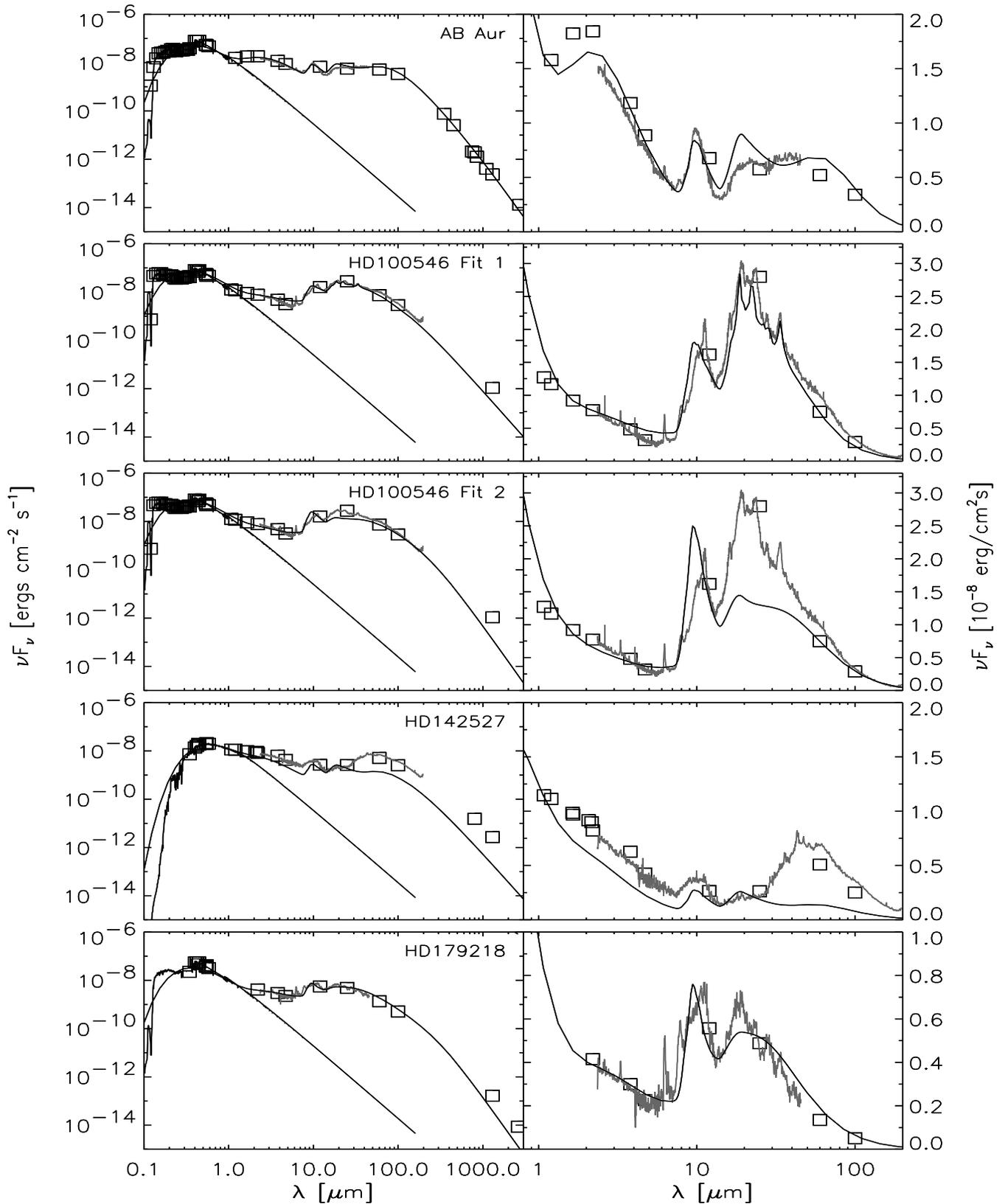}
\caption{The group Ia sources.  The left column shows the full SED
  $\nu F_{\nu}$ of
  the star on a log-log scale together with a Kurucz model for the
  star and a model fit for the disk.  The right column shows the
  region from 1 to 100 micron with a linear y-axis in order to allow
  for a better comparison between model and observations. The grey
  line show the ISO spectrum, the squares are photometry points.}
\label{fig:groupIa}
\end{figure*}
\begin{figure*}[t!]
\includegraphics[height=22cm,width=19cm]{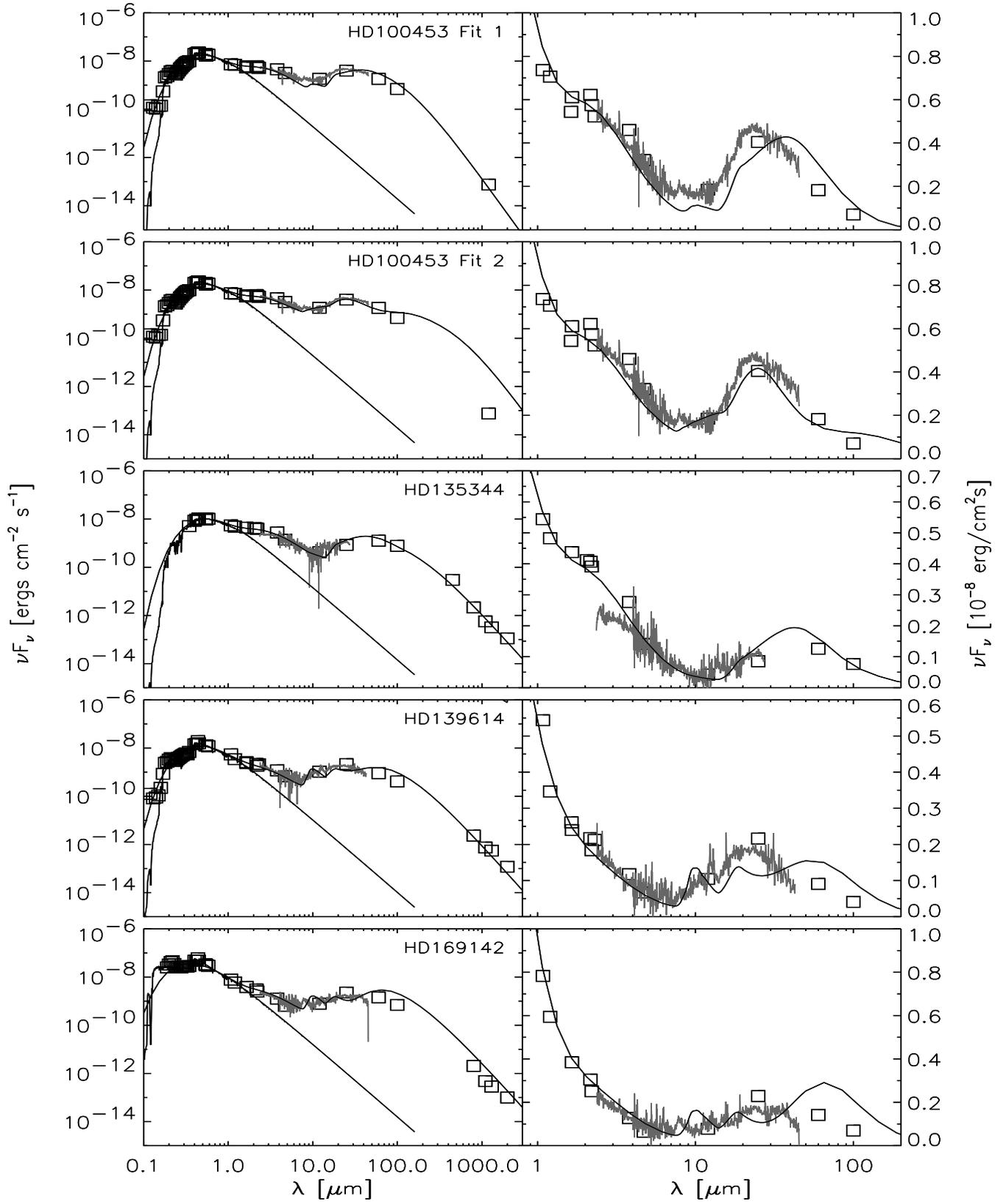}
\caption{The group Ib sources.  See caption of figure~\ref{fig:groupIa}.}
\label{fig:groupIb}
\vspace*{1cm}
\end{figure*}
\begin{figure*}[t!]
\includegraphics[height=22cm,width=19cm]{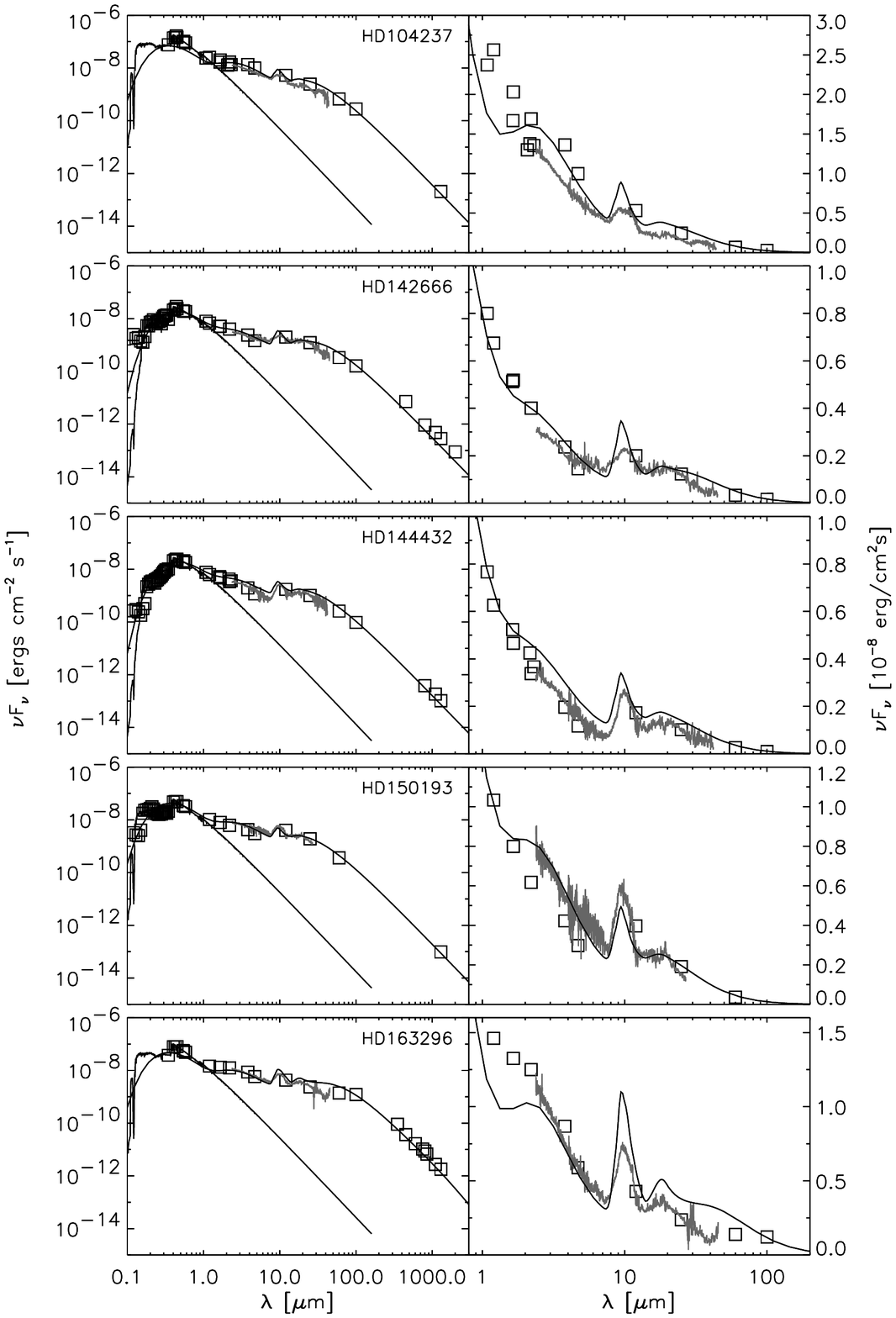}
\caption{The group ITa sources.  See also caption of figure~\ref{fig:groupIa}}
\label{fig:groupIIa}
\vspace*{1cm}
\end{figure*}


\section{Results}
\label{sec:results}
\subsection{The group Ia sources}

The group Ia sources show SEDs which are reminiscent of the SEDs of
T\,Tauri stars.  The SED is flat or even rising from 1 to about 60\um,
a clear sign for a flaring disk geometry.  Typically about 50\% of the
stellar radiation is reprocessed by the disk and emitted in the
infrared, approximately equally distributed between the 3\um{} bump between 2
and 7\um, and the rest of the infrared spectrum.  AB\,Aur is the
prototype of this group, and the best studied  Herbig Ae star.
Apart from AB\,Aur the Meeus et al.~sample contains three more stars in
the group Ia: HD\,100546, HD\,142527, and HD\,179218.  The SEDs and our
fits for these stars are shown in figure \ref{fig:groupIa}, and the
fit parameters are given in table \ref{tab:parameters}.  It is
immediately visible from the plots, that, while these sources are
similar to AB Aur, they also show pronounced differences.

\subsubsection{AB Aurigae as a prototype}

In the DDN paper the standard source was AB Aurigae, which is the best
studied Herbig Ae star. We show our fit to the SED in figure
\ref{fig:groupIa} and list the parameters used in table
\ref{tab:parameters}.  The parameters slightly differ from those used
by DDN.  The current values are more consistent with the measured size
of the disk and the inclination inferred from mm observations
\citep[e.g.][]{1997ApJ...490..792M}.  The model fits AB Aur very well. The
main difficulties are a small under-prediction at the onset of the
excess and an incorrect ratio of the 10\um{} and 20\um{} amorphous
silicate features.  The model parameters show what is typically needed
to fit a group I source: a decreasing surface density and an outer
radius of several hundred AU.  The submm slope is steep, so the normal
Draine \& Lee silicate opacities can be used to fit this source.

\subsubsection{HD 100546}

HD\,100546 is a very interesting group Ia source.  It shows an
extremely steep rise starting at 7\um{}.  The spectrum is dominated
by very strong emission features of crystalline silicates
\citep{1997Ap&SS.255...43M}.  In order to reproduce the crystalline
features, we introduced crystalline Forsterite as an additional dust
component for this star.

In AB Aur about 50\% of the reprocessed radiation is located in the
rim component and 50\% in the flaring component.  In HD 100546 these
corresponding numbers are 30\% and 70\% \citep{hd100546}.  With the
self-consistently computed inner rim height, the rim component becomes
much too luminous and the flaring component falls far below the
observed huge bump between 7 and 100 \um{}. The only solution which
produced a decent fit was to enforce a lower rim with $\chiin=1.2$
instead of the usual 4-5.  This reduces the rim component to the
observed values.  The flaring part of the disk then receives more
radiation and becomes more luminous, illustrating the trading of
energy between components in a passive disk model.  With a radially
decreasing surface density, the scale height of the disk in the
flaring region was still too small to reproduce the flux in the
20-40\um{} region, but a satisfactory result was achieved with a
\emph{constant} surface density.  In practice this means that most of
the disk material is located in the outer regions of the disk.  The
region between the rim and about 10 AU is almost empty.  This is
therefore similar to the gap proposed by \citet{2001A&A...375..950B}.
In order to produce the strong crystalline emission features, we need
a ratio of crystalline to amorphous material of 1:3, consistent with
the cool component in the Bouwman et al.~model.

A low surface density in the inner regions also provides a physical
explanation for the low rim height.  To illustrate this, we have also
computed an alternative model, in which we assume the surface density
to \emph{increase} with distance from the star.  We show this fit in
the third row of figure \ref{fig:groupIa}.  An increasing surface
density leads in practice to even less mass in the inner disk
regions which will then become marginally optically thin.  If we use
$p=2$, we can reproduce the low height of the inner rim in a
selfconsistent way.  However, the overall fit of this model is not quite as
good as the first model. In particular the ratio between 10 and 20
micron region becomes incorrect.

\subsubsection{HD 142527}
The fit to HD142527 is clearly much less satisfactory, but this is not
surprising.  The numbers listed in table~\ref{tab:measured} show that
the IR excess emission contains almost as much flux as the star
itself, a property which is impossible to fit with a passive
reprocessing disk model. Both the NIR and the FIR components are
larger than in any other star in the sample.  Reproducing the emission
in the 3 micron region would require a high inner rim, with
approximately 7 pressure scale heights, unrealistic for a hydrostatic
disk.  In addition, a high rim casts a long shadow on the flaring disk
region, so that the model necessarily predicts a relatively weak
far-IR spectrum, and gives hardly any rise at 20\um{}.  If indeed the
near-IR flux originates from the inner rim of the disk, then the huge
bump in the flux from the object around 60\um{} probably does not
originate from a disk, but from additional material in the same line
of sight.  Another possibility is that we see this star partially
through the disk, and that therefore the stellar flux is greatly
underestimated.  In fact, we have found several different estimates
for the luminosity of HD 142527, including one which is a factor of
two higher \citep{1997A&A...324L..33V}.  However, even with this
higher luminosity, no satisfactory fit could be obtained.  It is
possible that within the beam of both photometry and ISO
observations, there is another source of radiation peaking at 60\um.
However, the source is likely associated with HD\,142527 since the
emission shows significant crystalline silicate features implying
processed material, as well as crystalline $H_2O$ ice.

\subsubsection{HD 179218}
Somewhat similar to HD 100546, HD 179218 has a relatively low near-IR
emission, and a strong far-IR component.  Again we need an extreme
powerlaw for the surface density in this star: $p=2$. So also for this
star, most of the mass is located in the outer regions of the disk.
The resulting low rim height reproduces the low near-IR flux.  Most of
the reprocessing takes place in the flaring part of the disk, which
reproduces the mid- and far-IR flux. The width of the SED is much smaller
than in AB Aur, which is consistent with the small disk size we
derived.

\subsection{The group Ib sources}

The overall shape of spectra of the group Ib sources are similar to
the ones from group Ia, but very conpiciously the silicate feature is
absent in these spectra. There are two main possible explanations for
the absence of emission features.  Either the dust grains are all
large (a=3\um{} or larger), so that they will be opaque at 10\um{} and
not produce a feature.  Or the geometry of the disk is such that no
silicate grains of the right temperature are present.  DDN noted that
this can be achieved with a purely geometrical effect: If the inner
rim of the disk is high enough (about 5-7 pressure scale heights),
then the region of the disk which normally produces the 10\um{}
silicate emission feature is located entirely in the shadow, and the
corresponding component is missing from the spectrum. The
self-consistently computed rim height is usually not large enough to
achieve this.  We therefore tried to fit the SEDs of these sources by
treating $\chiin\equiv (H_s/H_p)_{\mathrm{rim}}$ as a free parameter.
Figure~\ref{fig:groupIb} shows our attempts to do so.  Similarly, a
gap in this 10\um{}-feature producing region could also have this
effect, but the possibility for a physical gap is currently not included
in the DDN model.

\subsubsection{HD 100453}

The ISO spectrum of HD\,100453 has the best signal-to-noise ratio of
all group Ib sources, and therefore provides the strongest constraints
on the absence of solid-state emission features
\citep{meeus-silicate}.  A very weak feature may be hidden in the
noise of the spectra.  In fact, van Boeckel et al. (2002, in
preparation) show with high quality ground-based data that even
HD\,100453 possesses a very weak feature. \newline In order to fit the
spectrum of HD100453, we used a rim height of 6 pressure scale
heights, as opposed to the computed 3.3.  This height is sufficient to
shadow the 10\um{} emitting region.  However, the large shadow also
decreases the emission from the disk interior at these radii, and the
model falls short between 8 and 30\um{}.  Our best model which is
shown in figure~\ref{fig:groupIb} shows a steep rise, but at longer
wavelengths than observed.  We have not been able to find a set of
parameters which at the same time suppresses the silicate feature and
fits the steep rise in the spectrum starting at 12\um{}.  At mm
wavelength, the fit is very good again, running through the 30mJy
measurement (S. Wolf, priv. communication).

The other possibility to produce a spectrum without a silicate
emission feature is to use large grains.  It has been shown by
\citet{meeus-silicate} that a depletion of hot small silicate grains
by a factor 300 compared to AB\,Aur would make the silicate feature
disappear.  In order to demonstrate this possibility, we have made a
second fit to the SED of HD 100453, using only large (3.3\um) silicate
grains.  This fit is shown in the second row of figure
\ref{fig:groupIb}.  Here we can fit the SED very satisfactory.
However, since the opacity per unit mass is reduced significantly, we
need an unphysically large disk mass (2 M$_{\odot}$) in order to keep
the required scale height of the disk, in particular in the rim
region.  The obvious solution to this problem would be to use a dust
composition which is not well mixed over the entire disk.  In
particular, if the opacity of the surface layer is provided by
non-silicate dust (e.g. small iron grains, carbon grains or even large
molecules like PAHs), the absence of the silicate feature could be
achieved by either coagulating the silicate grains to large ones, or
by settling the small silicates so that they do not receive direct
stellar radiation.

\subsubsection{HD 135344}

The fit for HD135344 is satisfactory from about 10 micron all the way down
into the mm region.  The inner rim height had to be increased 
to $\chiin=6$, and we need a relatively large disk (800 \AU)
in order to get the rather broad spectrum observed. In the region below
10\um{}, a significant difference between the photometry points and the ISO
spectrum can be observed.  The model fit is good for the SWS spectrum, but
clearly stays below the photometry points. The reason for this observational
discrepancy is not known.  Maybe the source is variable in the
NIR. Monitoring at these wavelengths would therefore be useful.

\subsubsection{HD 139614}

The model fit for HD139614 suffers from the same problems as the fit
for HD100453.  We can fit the NIR emission and the absence of the
silicate feature with an artificially high inner rim, but the price
for this is again too little flux in the 10-40\um{} region. Moreover,
HD139614 has a relatively low near-IR flux, which can only be
reconciled with the high inner rim when the inclination is very small
(face-on).  See HD169142 below for a discussion on this problem, which
is even more accute in that source. A reasonably over-all fit can be
found (see table \ref{tab:parameters}), but a still relatively strong
silicate feature appears.

\subsubsection{HD 169142}


The SED of the star HD 169142 is characterized by a low infrared
flux compared to the other members of its group. In terms of an optically
thick disk model this can only be interpreted in two ways: either we see the
disk at an inclination in which it happens to emit only very little flux, or
the disk itself has a very low reprocessing capability. In the purely
absorbing-reemitting disk model of DDN, this means that the disk covers a
relatively small fraction of the sky as seen by the star, i.e.~the disk is
rather flat. This `covering fraction', however, is a very weak function of
the disk parameters. Only by strongly reducing the outer radius of the disk,
and suppressing the height of the inner rim, one can achieve this
naturally. \citet{2001ApJ...547.1077C} suffered from a similar problem. They
suggested that dust settling has lowered the $\tau=1$ surface of the disk,
and thus the covering fraction of the disk is reduced.

We fitted a DDN disk with relatively low optical depth at large radii
(i.e. small mass, and steep $\Sigma$-powerlaw), and with a very small
inclination. The small inclination suppresses the near-infrared flux
because the emission from the inner rim is emitted as a cylinder
emitting from the inside: when seen at $i=0$, the flux is zero.  It is
unclear how realistic this is because of the uncertain geometrical
shape of the inner disk rim.  Furthermore, the best fit still shows a weak 10
micron silicate feature which seems to be absent in the spectrum.
Higher S/N ratio observations would be very useful to
confirm the presence or absence of a silicate feature in this star.

\subsection{The group IIa sources}

The overall spectrum of the group IIa sources show similarities and
differences with the group Ia sources. The initial onset of the IR
excess is still at 3\um{}, consistent with an inner disk boundary at
the evaporation temperature of the dust.  Also, the amount of energy
reprocessed in this component is similar.  However, the group II
sources don't show an additional rise of the spectrum beyond 20\um{},
so the flaring component is strongly suppressed. In the frame of the
DDN model this implies a truncation of the disk to very small sizes.

However, since the submm fluxes of theses sources are comparable to
the fluxes measured for group Ia sources, the disk mass in both cases
has to be similar. In view of the very small disk size, the disk is
therefore completely optically thick up to wavelengths of about 100 to
sometimes 1000 micron.  The only model for which the disk becomes
optically thin at 1000 micron is the model for HD150193.  At first
sight, this seems to be inconsistent with the steep slope of the
spectra of group II sources (with significantly less power in the
far-IR than in the mid-IR), since optically thick reprocessing disks
tend to produce mid- and far-IR fluxes of equal strength in $\nu
F_\nu$. However, due to the very small radius of the disk, the flaring
component is weak in comparison to the inner rim component.
The inner rim component will significantly contribute to the
mid-IR flux, but not to the far-IR flux. This naturally produces the
steep slope observed.  The truncation of the disk is such, that the
disk starts to flare just before the truncation radius.  The
small flaring is mainly required 
in order to produce a silicate emission feature which is present in all
known group II sources.  Meeus et al speculated that this may be a selection
effect: group II sources are relatively weak IR emitters, and
therefore only the brightest have been selected for the ISO study.

An alternative to a truncated disk is a collapsed disk outside of about
20AU, so that the material outside this radius is not directly exposed to
the stellar light, and will only be heated indirectly by the infrared
radiation of the inner part of the disk.  In the context of the simplified
model of DDN such a disk cannot be modeled properly.  Detailed 2-D radiative
transfer is required, which is outside the scope of this paper.
Clearly, such a model could help solve the difficulties with fitting
some of the far-IR an submm fluxes by supplying extra flux at these
long wavelengths.

\subsubsection{HD 104237}

This object is quite well fitted with a small disk. There is some freedom in
choosing the outer radius and the power law slope of the surface density, as
long as the outer radius is of order $10 \AU$ to $20 \AU$. For an outer
radius smaller than $3 \AU$ the entire disk would be in the shadow of the
inner rim, and we exclude this possibility. In the near-IR the photometry
points are considerably above the ISO points.  Out fits runs between
the two measurements.  A similar problem happens at around 20-100
microns.  Our fit matches the photometry points better than the ISO
fluxes, and runs straight through the 1.3 millimeter point.


\subsubsection{HD 142666}

Also for this star, a compact disk give a very good fit throughout the
IR region of the spectrum.  The model again underpredicts the mm
fluxes, so some cold dust, perhaps in a collapsed outer disk may be
needed to explain the entire spectrum.

\subsubsection{HD 144432}

A fully self-consistent (including $\chiin$) model fit can be obtained for a
small disk. This time, the mm points do not cause a problem for the
self-consistent determination of $\chiin$.

\subsubsection{HD 150193}

Also for HD150193 a fully self-consistent (including $\chiin$) model fit can
be obtained for a small disk.

\subsubsection{HD 163296}

The spectrum of HD163296 seems to be intermediate between group I and
group II sources in several ways.  The decrease in $\nu F_{\nu}$ only
starts at 100 micron, which is at longer wavelengths than for the
other group II sources.  Initially the disk SED starts to decrease at
about 30 micron like the other sources, but then recovers until 100
micron.  Such an extra ``bump'' cannot be fitted easily with a
continuous flaring disk.  Our fit is therefore a compromise: It tries
to fit the 100 micron point and consequently overpredicts the flux in
the 60 micron region.  The outer radius of the model fit is given by
50\,AU, also intermediate between the typical values for group Ia and
group IIa.  This size is still a factor of two smaller than the
continuum measurement of 100\,AU \citep{1997ApJ...490..792M}, while
the CO data even gives 310\,AU.

\section{Discussion}
\label{sec:discussion}

The attempt to fit the spectra of isolated Herbig Ae stars with a
physical disk model requires different typical sets of parameters for
the different disks.  Overall, the fits are reasonable for group Ia
and even better for IIa.  Group Ib sources require large grains to
hide the silicate feature effectively without affecting the cold parts
of the disk.  What remains to be seen is if the fit parameters are
physically realistic and reasonable.

\subsection{Group Ia}

The majority of the goup I sources can be well fitted with large (few
hundred AU) disks.  The surface density distribution needed for these
sources varies greatly.  For AB Aur, a powerlaw $p=-1.5$ works.  This
surface density slope is consistent with the structure of the solar
nebula as derived from the planet masses, and also a reasonable result
for accretion disks \citep{1981PThPS..70...35H}.  On the other hand,
for HD\,100546 and HD\,179218 we find that a surface density
increasing with distance from the star is required.  This can be
understood as a largely cleared-out inner region with still copious
amounts of material present in the outer regions of the disk.  This
picture is quite different from a normal accretion disk which leads to
a decreasing surface density with distance.  In disks with gaps
(possibly created by planets), accretion accross the gap will be
severely decreased \citep{1996ApJ...467L..77A} and the supply of
material which feeds the inner disk dies out.  Since the viscous
timescales in the inner disk are short, the entire part inside the gap
will be emptied quickly, which may produce a structure consistent with
the density distribution required for our fits.  We suggest that the
inner disks of these two sources are starved, and have lost much of
their initial mass content by accretion.  The current accretion rate
in these sources should be extremely low.

\subsection{Group Ib}

The attempt to explain the absence of the silicate feature in the
group Ib sources by rim shadowing seems to have failed.  First of all,
the required rim height is considerably higher than typical
hydrostatic rims can provide.  The height of the rim is defined as the
highest radial ray through the inner rim which still becomes
optically thick for stellar radiation.  In hydrostatic equilibrium,
the vertical density structure in the disk is given by a Gaussian
distribution.  For normal surface densities, the surface height in the
rim is typically between 4 and 5 pressure scale heights, clearly in the
exponential range of the Gaussian density distribution.  In order to
change the densities enough for a 30\% increase of the rim height, the
surface density in this region would have to increase by a factor of 10.
Therefore, a hydrostatic rim height of 6 or even 7 pressure scale
heights seems unrealistic.  It may be possible to have hydrodynamic
fluctuations sling matter towards higher elevations above the
midplane, a phenomenon which might be related to the variability
observed in UX Orionis stars.
However, such a scenario would predict a variable shadowing effect
rather than a complete blocking of the stellar light.  Due to the lack
of spectra taken at different epochs, such a variability cannot be
excluded at the present time.

Another possible way to suppress the 10 micron feature by a
geometrical effect is by ``frustrated flaring''. In the DDN picture,
the disk rises out of the shadow as soon as this is possible, assuming
the shape of a normal Chiang \& Goldreich flaring disk in the outer
regions.  However, it cannot be ruled out that the flaring of the disk
starts further out in the disk.  It is then possible that the 10
micron emitting region remains in the shadow of a normal rim.  As the
disk reappears from the shadow at larger radii, its flaring will be
stronger than usual to compensate for the lost flux.  This would
produce a larger-than-normal bump in the spectrum, of which the SEDs
of HD 100546, and HD 179218 may be examples.  In a recent study,
\citet{2002ApJ...568.1008C} have used a similar model to explain the
SED of the T Tauri star TW-Hydra.  They propose that a planet clears
out part of the inner region, so that the disk can only start to flare
further away from the star.

Finally, the best solution to the ``missing silicate mystery'' may be
that the silicate grains have coagulated to large grains or settled to
the disk interior, or both.  The opacity of the disk surface must in
this case be provided by other small grains or large molecules which
will neither settle nor coagulate, but still provide optical and UV
opacity.  The fact that all group IIa sources show PAH emission may
point into this direction. 

\subsection{Group IIa}

The group II sources can be fit really well with very compact disks
with outer radii as small as 10 AU.  Since the sub-mm measurements of
these disks are not very different from the group Ia sources, the disk
mass must be similar.  This combination of high mass and small size
leads to very high surface densities, and to disks which are optically
thick up to mm wavelengths.   Unfortunately, sizes for most of these
disk are unknown, with the only exception of HD\,163296, which is the
least compact disk according to our fits.  The observed size of
100\,AU is a factor of w larger that our best fit model.  

The compact size seems to be inconsistent with disk sizes typically
measured around young stars.  Our own solar system, the diameters of
some Vega-like debris disks \citep{1984ApJ...278L..23A} and the sizes
of proplyds observed in the Orion starforming region
\citep{1998AJ....115..263O} all point to disk sizes which are at least
50 AU and reach up to 1000 AU.

The unshadowed part of the flaring disk reprocesses less radiation
than the disk's inner rim, so that the spectrum has a strong 3 micron
bump and falls off (in $\nu F_\nu$) as one goes to longer wavelengths.
This conclusion is very similar to the conclusion drawn by
\citet{2001A&A...365..476M} that the group IIa sources have no flaring
disk.  But there is an important difference.  Our models have very
small flaring disks, but they are certainly not vanishingly small. The
10 micron feature is still produced entirely by the compact flaring
disk itself. Only the long wavelength part is suppressed by truncating
the disk at small radii.

What is suspect about these fits is the fact, that we need a very specific
location at which to truncate the disk.  For the production of the silicate
emission features, the disk needs to flare out of the shadow a little bit,
but we cannot allow the disk to reach beyond 20\,AU because then the
corresponding surface layer and midplane components would lead to a group
I-like SED.  What is truncating these disks at exactly the right
radius?  In particular, what is responsible for the striking
differences between group Ia and group IIa?  The near IR region as
well as the submm fluxes of both groups are similar, the distribution
on the sky and the absence of a nearby star forming region is the same
for both.  Why do the group IIa disk either collapse or shrink
(without loosing mass) to very small sizes?  We conclude that while
the DDN model is capable of fitting the SEDs of group IIa sources very
well, the model parameters derived seem difficult to understand.


\section{Summary}

In this paper we investigated the SEDs of the sample of 13 isolated
Herbig Ae stars of \citet{2001A&A...365..476M} in terms of a passive
irradiated flaring disk model.  We find that for most of the stars the
general shape of the SEDs can be naturally reproduced by the model.
This adds evidence to the idea that the infrared excess of these stars
is produced by a circumstellar disk, without need for additional
circumstellar components. The conspicuous 3-micron bump in these SEDs,
which until recently was considered an enigma, can be naturally
explained as the emission from the inner rim of these disks. All of
these disks have an inner rim emitting at 1500 K due to the fact that
dust evaporation removes the dust (and therefore the opacity) at radii
smaller than about half an AU. The disk picture is therefore a natural
explanation for the infrared excess of most of these stars.

The three different types of spectra identified by Meeus et al. can,
to a certain extent, be interpreted in terms of disk geometry. Their
group Ia sources can be explained by large disks, some of which have
cleared-out inner regions. The group IIa sources can be well fitted by
compact disks with outer radii in the order of 10 to 50 AU.  However,
it is not clear what could cause such small disk radii in combination
with the high surface densities needed to explain the disk mass.  A
collapsed outer disk which is not flaring may therefore be a more
attractive scenario, which we will be exploring in a future
publication.  Only the group Ib sources, in which the 10 micron
silicate feature is completely absent, seem to defy explanation in
geometrical terms only.  However, it is possible that additional
effects like grain settling and disk collapse may help to resolve
these differences.

\begin{acknowledgements}
  We would like to thank A. Natta for stimulating discussions related
  to this paper, partly during a very pleasant working visit of CD to
  Arcetri. 
  We would like to thank G. Meeus and M. van den Ancker for providing
  us with electronic version of their observed and dereddened data.
  CD and RW acknowledge the financial support from NWO Pionier grant
  6000-78-333.  CPD acknowledges support from the European Commission
  under TMR grant ERBFMRX-CT98-0195 (`Accretion onto black holes,
  compact objects and prototars').
\end{acknowledgements}


\end{document}